\documentclass{aa}

\usepackage{graphicx}
\usepackage{txfonts,rotating}
\begin{document}

\title{First albedo determination of 2867 Steins, target of the Rosetta mission
\thanks{Based on observations carried out at European Southern Observatory (ESO), Paranal, Chile, Prog. 075.C-0201(A).}}

\author{Fornasier S. \inst{1} \and
        Belskaya I. \inst{2} \and
        Fulchignoni M. \inst{3}  \and
	Barucci M.A \inst{3}  \and
        Barbieri C.\inst{1} 
        }

\offprints{S. Fornasier}

\institute{Astronomy Department and CISAS, Padova University, Italy \email{fornasier@pd.astro.it} 
\and Institute of Astronomy of Kharkiv National University, Ukraine
\and Paris Observatory, France 
}

\date{Submitted to A\&A on 25 January 2006, accepted on 9 February 2006}

%{\bf Running head}: First albedo determination of 2867 Steins

\abstract{}{
We present the first albedo determination of 2867 Steins, the asteroid target of the Rosetta space mission together with 21 Lutetia.}
{The data were obtained in polarimetric mode at the ESO-VLT telescope with the FORS1 instrument in the V and R filters. Observations were carried out from June to August 2005 covering the phase angle range from 10.3$^{o}$ to 28.3$^{o}$, allowing the determination of the asteroid albedo by the well known experimental relationship between the albedo and the slope of the polarimetric curve at the inversion angle. } 
{The measured polarization values of Steins are small, confirming an E-type classification for this asteroid, as already suggested from its spectral properties. The inversion angle of the polarization curve in the V and R filters is respectively of 17.3$\pm$1.5$^{o}$ and 18.4$\pm$1.0$^{o}$, and the corresponding slope parameter is of 0.037$\pm$0.003\%/deg and 0.032$\pm$0.003\%/deg. 
On the basis of its polarimetric slope value, we have derived an albedo of 0.45$\pm$0.1, that gives an estimated diameter of 4.6 km, assuming an absolute V magnitude of 13.18 mag.}{}


\maketitle

\section{Introduction}

Rosetta is the ESA cornerstone mission devoted to the study of minor bodies.
%After the launch postponement in January 2003, the orbit and the targets of the %Rosetta mission have
%been completely changed.  
Successfully launched on March 2, 2004, Rosetta, in its journey to the comet 67P Churyumov-Gerasimenko, will fly by two main belt asteroids, 2867 Steins on September 2008 and 21 Lutetia on July 2010. \\ 
All the targets have been changed with respect to the original mission plans due to the Rosetta launch postponement of about one year. Several observational campaigns have been promoted to increase knowledge of the physical properties of these new targets. \\ 
While 21 Lutetia is a large main belt asteroid with a long history of investigation, 2867 Steins has been observed in detail only from 2004, and its physical properties are still not completely known. \\
The first Steins lightcurve was reported by Hicks and Bauer (2004), who found a rotational period of 6.06$\pm$0.05 hours and an amplitude of approximately 0.2 mag, values confirmed by Weissman et al. (2005). \\
The spectroscopic investigation in the visible and near infrared range presented by Barucci et al. (2005) show a strong feature at 0.5 $\mu$m, a weaker feature at 0.96 $\mu$m and a flat and featureless behavior above 1 $\mu$m. This spectral behavior is very similar to that of the E--type objects (Fornasier \& Lazzarin 2001, Burbine et al. 1998), and in particular to that of the E subtype II (Angelina like) following Clark et al. (2004) and Gaffey \& Kelley (2004) classification scheme. Nevertheless, as E type objects are characterized by the highest albedo values found in the asteroid population, knowledge of Steins' albedo is needed for a definitive taxonomic classification and a proper understanding of its surface properties and size determination. \\
In this paper, we present the first polarimetric investigation of 2867 Steins and discuss its albedo and size evaluation based on polarimetric properties. This information will aid the Rosetta mission in planning science operations and optimizing the flyby trajectory.

\begin{table*}[t]
   \caption{Observational conditions for 2867 Steins. The exposure time (exp) is the same for each of the four images obtained at the four different $\lambda/2$ retarder plate positions, while nset represents the number of repetitions of the full polarimetric sequences. Data on the position angle of the scattering plane ($\phi$), on the phase angle ($\alpha$), on the heliocentric ($r$) and geocentric ($\Delta$) distances of Steins have been taken from the JPL ephemeris service (http://ssd.jpl.nasa.gov/cgi-bin/eph).}
\begin{center}
\label{tab1}
\begin{tabular}{|l|c|c|c|c|c|c|c|c|l|l|c|} \hline
DATE & UT & Fil. &  exp (s) & nset  & m$_{v}$ & airm & $\phi$ ($^{o}$)  & $\alpha$ ($^{o}$) & $r$ (AU) & $\Delta$ (AU) & seeing (") \\ \hline
11 Jun. 05 & 08:50 & R & 33 & 3 & 16.45 & 1.05 & 252.6 & 28.3 & 2.0201 & 1.4315 & 1.45 \\
11 Jun. 05 & 09:08 & V & 35 & 3 & 16.45 & 1.02 & 252.6 & 28.3 & 2.0201 & 1.4315 & 1.35 \\
01 Jul. 05 & 09:23 & R & 26 & 3 & 16.06 & 1.02 & 258.2 & 24.3 & 2.0197 & 1.2563 & 0.55 \\
01 Jul. 05 & 09:40 & V & 26 & 3 & 16.06 & 1.03 & 258.2 & 24.3 & 2.0197 & 1.2563 & 0.54 \\
14 Jul. 05 & 05:43 & R & 15 & 3 & 15.78 & 1.17 & 265.0 & 20.6 & 2.0217 & 1.1653 & 0.80 \\
14 Jul. 05 & 05:58 & V & 15 & 4 & 15.78 & 1.12 & 265.0 & 20.6 & 2.0217 & 1.1653 & 0.99 \\
06 Aug. 05 & 04:16 & R & 15 & 3 & 15.31 & 1.13 & 294.1 & 12.3 & 2.0295 & 1.0654 & 0.60 \\
06 Aug. 05 & 04:31 & V & 15 & 4 & 15.31 & 1.10 & 294.2 & 12.3 & 2.0295 & 1.0654 & 0.55 \\
09 Aug. 05 & 07:56 & R & 10 & 4 & 15.26 & 1.09 & 301.7 & 11.3 & 2.0310 & 1.0594 & 0.74 \\
09 Aug. 05 & 08:14 & V & 10 & 4 & 15.26 & 1.12 & 301.7 & 11.3 & 2.0310 & 1.0594 & 0.71 \\
13 Aug. 05 & 07:49 & R & 10 & 4 & 15.22 & 1.10 & 313.2 & 10.3 & 2.0330 & 1.0547 & 2.10 \\
13 Aug. 05 & 08:16 & V & 10 & 4 & 15.22 & 1.16 & 313.2 & 10.3 & 2.0330 & 1.0547 & 1.94  \\
\hline
      \end{tabular}
       \end{center}
       \end{table*}

\section{Data acquisition and reduction}

Observations of 2867 Steins were carried out in service mode at the VLT telescope UT2 KUEYEN with the FORS1 instrument in polarimetric mode (see http://www.eso.org/instruments/fors1) from June to August 2005. Linear polarimetry was obtained in the two broadband filters R and V at 4 angles of the $\lambda/2$ retarder plate (0$^{o}$, 22.5$^{o}$, 45$^{o}$ and 67.5$^{o}$ with respect to celestial coordinate system), covering the phase angle range from 10.3$^{o}$ to 28.3$^{o}$. \\
The procedure during each observing run included the acquisition of
several flat field images, taken during twilight time without polarimetric optics in the light path, and of at least one unpolarized standard star to calibrate the instrumental polarization.
The zero points of the position angles both for V and R images were taken from the FORS1 user manual (http://www.eso.org/instruments/fors/doc).  

The asteroid and standard star images have been corrected for bias and master flat, and the cosmic rays removed.
The center of the asteroid image in each of the two channels is
evaluated by a
2 dimensional centroid algorithm.  The flux for each channel is
integrated over
a radius corresponding to 3-4 times the average seeing, and
the sky is subtracted
using a 3-5 pixel wide annulus around the object. 

 The Stokes parameters are derived in the following manner: 
\begin{equation}
  Q = 0.5 \left[\left(\frac{f^{o}-f^{e}}{f^{o}+f^{e}}\right)_{\lambda/2 pos. = 0^{o}} 
-\left(\frac{f^{o}-f^{e}}{f^{o}+f^{e}}\right)_{\lambda/2 pos. = 45^{o}}\right]
\end{equation}

and
\begin{equation}
U = 0.5 \left[\left(\frac{f^{o}-f^{e}}{f^{o}+f^{e}}\right)_{\lambda/2 pos. = 22.5^{o}} 
-\left(\frac{f^{o}-f^{e}}{f^{o}+f^{e}}\right)_{\lambda/2 pos. = 67.5^{o}}\right]
\end{equation}
where f$^{o}$ and f$^{e}$ are the background subtracted object flux of the ordinary and extra-ordinary beam inside each image and $\lambda/2$ pos. indicates the position of the retarder plate. \\

The degree of polarization $P$ and the position angle $\theta$ of the polarization plane 
in the instrumental reference system are expressed via the parameters Q and U with the well-known formulae
\begin{equation}
 P = \sqrt{U^2+Q^2} \hspace{1cm} \sigma_{P} = \frac{|U*dU+Q*dQ|}{P}
\end{equation}
\begin{equation}
 \theta = \frac{1}{2} arctan\frac{U}{Q}  \hspace{1cm} \sigma_{\theta} = \frac{28.65*\sigma_{P}}{P}
\end{equation}
\noindent
where dU and dQ are the errors on the Stokes parameters (Shakhovskoy \& Efimov, 1972). \\
The position angle of the polarization plane relative to the
plane perpendicular to the scattering plane, $\theta_r$, is expressed as $\theta_r = \theta -  (\phi \pm 90^{o}) $, where $\phi$ is the position angle of the scattering plane and the sign 
inside the bracket is chosen to assure the condition $0^o \le(\phi \pm 90^{o}) \le 180^o$. \\
The polarization quantity $P_r$ is computed as $P_r = P * cos(2\theta_r) $. \\
During each run, Steins observations at the four retarder positions were repeated at least three times (see nset column in Table~\ref{tab1}) and the polarization values derived as the median of these multiple exposures.

\section{Results}
\begin{table*}
   \caption{Results of 2867 Steins polarimetric observations in the R and V filters.}
\begin{center}
\label{tab2}
\begin{tabular}{|l|c|c|c|c|c|} \hline
DATE & Fil. & P (\%) & $\theta$ ($^{o}$) & P$_r$ (\%) & $\theta_r$ ($^{o}$)  \\ \hline
11 Jun.  & R & 0.287$\pm$0.062 & 163.9$\pm$6.1 & +0.287$\pm$0.062 & 1.3$\pm$6.1  \\
11 Jun.  & V & 0.369$\pm$0.068 & 171.7$\pm$5.2 & +0.351$\pm$0.062 & 9.1$\pm$5.2  \\
01 Jul.  & R & 0.204$\pm$0.052 & 170.0$\pm$5.2 & +0.204$\pm$0.054 & 1.8$\pm$5.2  \\
01 Jul.  & V & 0.309$\pm$0.047 & 166.5$\pm$4.3 & +0.309$\pm$0.050 & -1.7$\pm$4.3  \\
14 Jul.  & R & 0.178$\pm$0.057 & 25.6$\pm$9.7 & +0.086$\pm$0.057 & 30.6$\pm$9.7  \\
14 Jul.  & V & 0.146$\pm$0.060 & 11.1$\pm$11.7 & +0.124$\pm$0.060 & 16.2$\pm$11.7  \\
06 Aug.  & R & 0.209$\pm$0.048 & 113.9$\pm$6.5 & -0.209$\pm$0.050 & 89.8$\pm$6.5  \\
06 Aug.  & V & 0.205$\pm$0.053 & 114.5$\pm$6.8 & -0.204$\pm$0.053 & 90.4$\pm$6.8  \\
09 Aug.  & R & 0.213$\pm$0.068 & 122.3$\pm$9.1 & -0.213$\pm$0.068 & 90.6$\pm$9.1  \\
09 Aug.  & V & 0.284$\pm$0.080 & 155.1$\pm$10.9 & -0.205$\pm$0.080 & 111.9$\pm$10.9  \\
13 Aug.  & R & 0.265$\pm$0.071 & 140.7$\pm$7.5 & -0.256$\pm$0.071 & 97.5$\pm$7.5  \\
13 Aug.  & V & 0.239$\pm$0.080 & 135.0$\pm$9.5 & -0.238$\pm$0.080 & 91.8$\pm$9.5  \\
\hline
      \end{tabular}
       \end{center}
       \end{table*}
The polarimetric characteristics of 2867 Steins derived from the above observations are reported in Table~\ref{tab2} and visualized in Fig. 1. 
Our data permit the determination of the inversion angle and of the slope of the polarization phase curve at 
the inversion angle in the V and R bands. We used the linear fit to the data weighted according to their errors (see Fig. 1). The use of a linear fit is reasonable because of the small curvature of the ascending branch of the asteroid polarization phase dependence (Zellner \& Gradie 1976). The obtained parameters are given in Table~\ref{tab3}. 

For comparison, in Table~\ref{tab3} we also give the mean values of the corresponding parameters for the main asteroid types according to Goidet-Devel et al. (1995). The polarimetric characteristics of Steins correspond to the mean values of E-type asteroids. \\
Currently, detailed polarimetric observations are available only for two E-type asteroids, 44 Nysa and 64 Angelina. We report in Fig. 2 observations in the V band obtained with an accuracy better than 0.1\% together with our data on Steins, showing that all the three asteroids have very similar polarimetric properties. 
%data are in very well agreement with each other. 
Fitting the data with the linear-exponential function (Fig. 2) proposed by Kaasalainen et al. (2003) gives almost the same polarimetric slope (0.038$\pm$0.006\%/deg) derived by a linear fit.

\begin{figure}
   \centering
    \includegraphics[width=9.5cm]{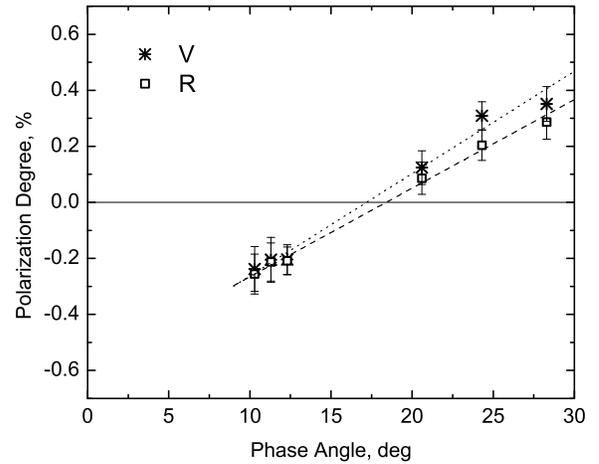}
   \caption{Polarization degree versus phase angle in the V and R bands for asteroid 2867 Steins.}
              \label{fig1}
    \end{figure}

\begin{figure}
   \centering
    \includegraphics[width=9.5cm]{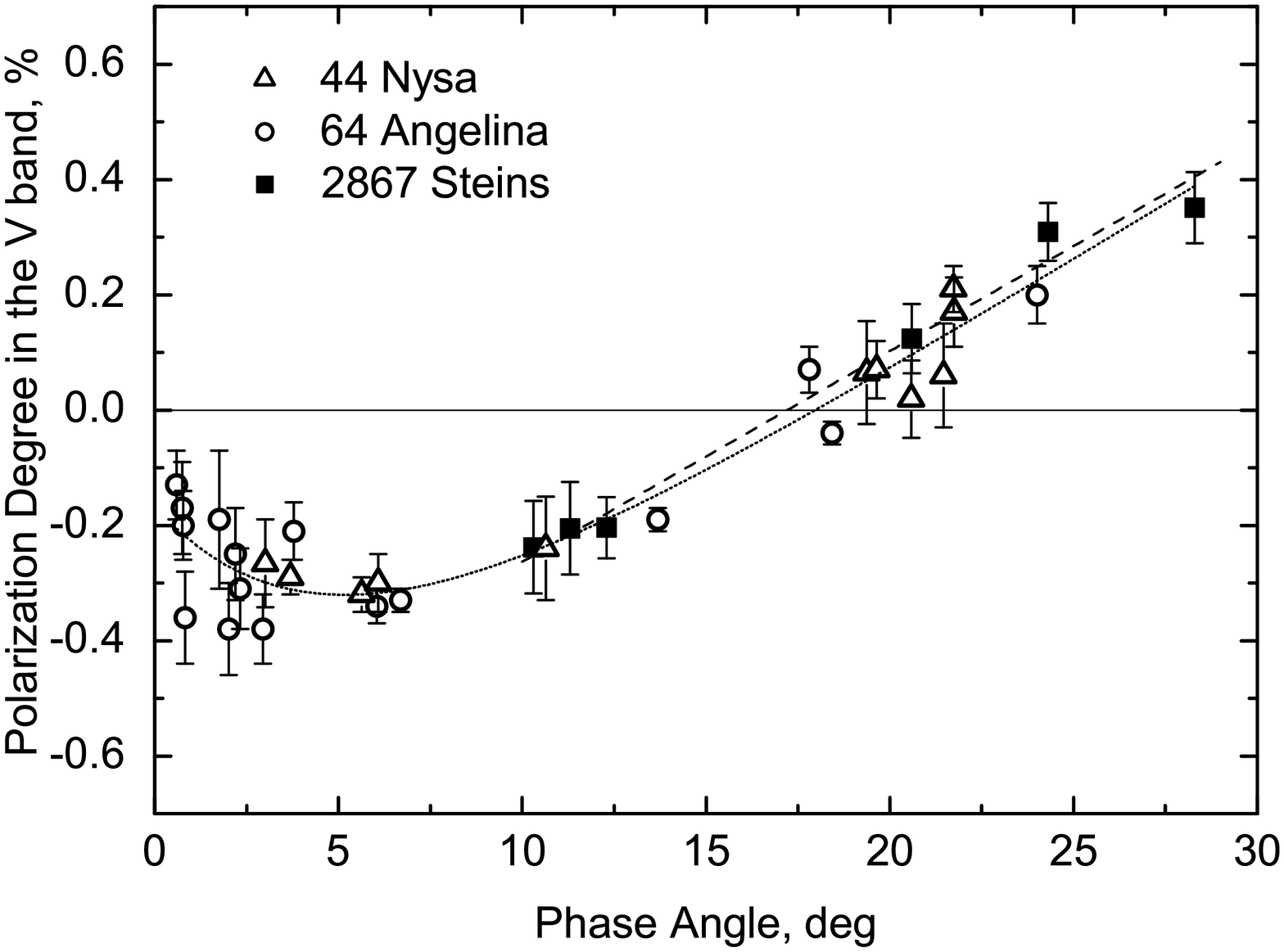}
   \caption{Polarization - phase dependence of E-type asteroids. 
   Data for 44 Nysa and 64 Angelina are taken from Zellner \& Gradie (1976), Cellino et al. (2005),
   Rosenbush et al. (2005), Fornasier et al. (2006). 
   The continuous line shows the fitting of all data by the linear-exponential function proposed by Kaasalainen et al. (2003); the dashed line shows the linear fit to Steins' data. }
              \label{fig2}
    \end{figure}

\begin{table}[t]
   \caption{Polarimetric parameters of 2867 Steins in comparison to mean values for the main asteroid classes (Goidet-Devel et al. 1995).}
\begin{center}
\label{tab3}
\begin{tabular}{lll} 
AST. & slope (\%/deg) & inv. angle ($^{o}$) \\ \hline \hline
2867 Steins    &     0.037$\pm$0.003 (V)  & 17.3$\pm$1.5 \\
               &     0.032$\pm$0.003 (R) & 18.4$\pm$1.0 \\
E-type         &     0.04             & 17.8        \\
S-type         &     0.09             & 20.1        \\
M-type         &     0.09             & 23.5        \\
C-type         &     0.28             & 20.5        \\ \hline
    \end{tabular}
       \end{center}
       \end{table}

As it is well known, accurate measurements of the polarimetric slope are very important for the determination of the asteroid albedo. The empirical correlation of polarimetric slope vs. albedo has been successfully used for asteroid albedo determinations e.g. by Zellner \& Gradie (1976) and Cellino et al
. (1999, 2005). It has the simple form: 
\[ \log (p_v) = C1\times\log(h) + C2 \]  
where p$_{v}$ is the geometric albedo and C1 and C2 are empirical constants. The values of these constants are slightly different, depending on the dataset used for their determination. The constants used by Zellner \& Gradie (1976) were derived from laboratory data for meteorites and terrestrial samples (Bowell \& Zellner 1974). Lupishko \& Mohamed (1996) gave new constants based on asteroid data using albedos derived from different sources, including IRAS data, Earth-based radiometric observations and stellar occultation.  Cellino et al. (1999) derived the constants from the data set of asteroids with well-measured IRAS albedos. \\
The differences between the albedos calculated using different constant values are shown in Fig. 3. 
The main discrepancy arises for high albedo asteroids due to the very few objects for which data are available. 
The measured polarimetric slope of Steins in the V band is very similar to that of 64 Angelina (see Fig. 3). \\
\begin{figure}[t]
   \centering
    \includegraphics[width=9.5cm]{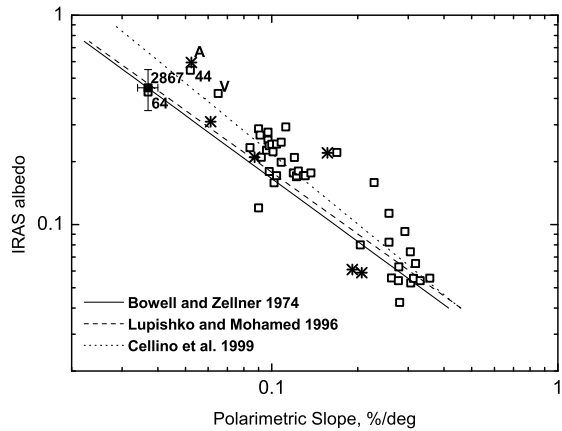}
   \caption{IRAS albedo versus polarimetric slope. Polarimetric data taken from Lupishko \& Mohamed (1996) are marked with open squares and data from Fornasier et al. (2006) with star symbols.  Lines represent albedo--slope dependences used by different authors for albedo calculations. High--albedo asteroids are marked by their spectral types (V and A). E--type asteroids are designated by number. }
              \label{fig3}
    \end{figure}
Unfortunately, Angelina's albedo has not been measured from the IRAS satellite. The usually adopted albedo value of 0.43 was given by Tedesco et al. (1989), with 
reference to unpublished Earth-based radiometric observations.  
Later Tedesco et al. (2002) derived a value of 0.40$\pm$0.05 from the Midcourse Space Experiment data.
At the same time, the IRAS albedo of another E-type asteroid, 44 Nysa, is higher ($p_{v} = 0.55$), in spite of the larger polarimetric slope (see Fig.3). Its polarimetric properties are similar to those of the A-type asteroid 863 Benkoela (Cellino et al. 2005; Fornasier et al. 2006), which has an IRAS albedo of 0.6. However, such high values of IRAS albedos could be an overestimation connected with using non-simultaneous photometric and radiometric data.  
Uncertainty in absolute magnitude is more critical for high albedo asteroids, as shown by Harris \& Harris (1997). Thus, to estimate Steins' albedo we prefer using constants derived  by Bowell \& Zellner (1974) from laboratory data. They give an albedo estimation independent of other techniques. 

The albedo of Steins based on its polarimetric slope value is 0.45$\pm$0.1 (see Fig.3), where the error comprises both the uncertainty on the slope and that of constants in the slope-albedo relation. Assuming an absolute visual magnitude $V = 13.18$ mag (Hicks et al. 2004), its estimated diameter is approximately 4.6 km.

Our observations also show systematic differences in the V and R bands at phase angles larger than the 
inversion angle (Fig.1). Both positive polarization degree and polarimetric slope decrease with increasing 
wavelength, while negative polarization is practically the same in both filters. This behavior reflects the inverse 
dependence of positive polarization on albedo observed also for S-type asteroids (e.g. Lupishko et al. 1995).\\

\section{Conclusions}

Our polarimetric investigation shows that 2867 Steins is a high albedo asteroid (p$_{v}$ = 0.45$\pm$0.1) with an estimated diameter of approximately 4.6 km.    
The high albedo, together with the peculiar polarimetric properties typical of E-type asteroids, lead to the conclusion that 2867 Steins is an E-type object, as already suggested on the basis of its spectral behavior (Barucci et al., 2005). \\
 Rosetta will be the first space mission to fly by an E-type asteroid, and its results will be very important in the understanding of the physical properties of this peculiar class, whose high albedo members show different spectral behaviors (Clark et al., 2004).

\begin{acknowledgements}
This work has been partly supported by the ASI-Rosetta contract to CISAS (University of Padova) and by an Europlanet personnel exchange grant.
\end{acknowledgements}


\begin{thebibliography}{}
\bibitem[2005]{bar05} Barucci M. A., Fulchignoni M., Fornasier S., et al. 2005. \aap, 430, 313--317.
\bibitem[1974]{bow74} Bowell, E. \& Zellner B. 1974. In: {\it Planets, stars, and nebulae studied with photopolarimetry}, edited by T. Gehrels, Univ. Arizona Press, Tuscon, p. 381.
\bibitem[2004]{clark04}  Clark B. E., Bus S.J., Rivkin A.S., et al. 2004. \jgr, 109, 2001--2012.
\bibitem[2005]{cellino05} Cellino, A., Hutton, R. G., di Martino, M., et al. 2005. Icarus, 179, 304--324.
\bibitem[1999]{cellino99} Cellino, A., Hutton, R. G., Tedesco, E. F., di Martino, M., Brunini, A., 1999. Icarus, 138, 129-140.
\bibitem[2006]{fornasier06} Fornasier, S., Belskaya, I., Shkuratov, Yu. G., et al., 2006. Submitted to \aap.  
%\bibitem[2001]{fornasier01} Fornasier, S., Barucci M.A., Binzel %R.P. et al. 2003. \aap, 398, 327--333.  
\bibitem[2001]{fornasier01} Fornasier, S. \& Lazzarin, M., 2001. Icarus, 152, 127--133.
\bibitem[2004]{Gaffey04} Gaffey, M.J. and Kelley, M.S. 2004. LPI XXXV, 1812
\bibitem[1995]{Goidet95} Goidet-Devel, B., Renard, J.B., Levasseur-Regourd, A.C., 1995. \planss, 46, 779--786.
\bibitem[1997]{harris97} Harris A.W. \& Harris, A.W., 1997. Icarus, 126, 450--454.
\bibitem[2004]{hicks04} Hicks,M.D., Bauer,J.M., Tokunaga,A.T. 2004. \iaucirc, 8315, 3. 
\bibitem[2003]{Kaasalainen03} Kaasalainen, S., Piironen, J., Kaasalainen, M., et al., 2003. Icarus, 161, 34--46.
%\bibitem[2001]{lebras2001} Le Bras, A., Dotto, E., Fulchignoni %M., et al., 2001, \aap, 379, 660--663. 
\bibitem[1996]{lupishko96} Lupishko, D. F. \&  Mohamed, R. A., 1996. Icarus, 119, 209-213.
\bibitem[1995]{lupishko95} Lupishko, D. F., Vasilyev, S. V., Efimov, Yu.S.,  Shakhovskoj, N. M., 1995. Icarus, 113, 200--205.  
\bibitem[2005]{rosen05} Rosenbush, V. K., Kiselev, N. N., Shevchenko, V. G., et al., 2005. Icarus, 178, 222--234.
\bibitem[1972]{sh72} Shakhovskoy N. M. \& Efimov, Yu. S., 1972. Izv. Krymskoi Astrofiz. Obs., 45, 90--110.
\bibitem[1989]{tedesco89} Tedesco, E. F., Williams, J. C., Matson, D. L., et al., 1989. In: {\it Asteroids II}, edited by Binzel, R.P.,  Gehrels T. and Matthews, M.S., Univ. of Arizona Press, Tucson,  p. 1151--1161.
\bibitem[2002]{tedesco02} Tedesco, E. F., Egan, M. P., Price, S. D. 2002. \aj, 124, 583--591.
\bibitem[2005]{weiss05} Weissman, P. R., Lowry, S. C., Choi, Y. J., 2005. American Astronomical Society, DPS meeting 37, 15.28
\bibitem[1976]{zellner76} Zellner, B. \&  Gradie J., 1976. \aj, 81, 262--280.


 
\end{thebibliography}
\end{document}